\documentclass[aps,pra,10pt,showpacs,amsmath,longbibliography,twocolumn,floatfix,eqsecnum]{revtex4-1}
\usepackage[dvipsnames]{xcolor}
\usepackage[colorlinks=true,linkcolor=NavyBlue,urlcolor=NavyBlue,citecolor=NavyBlue,breaklinks]{hyperref}
\usepackage{amsmath}
\usepackage{amsfonts}
\usepackage{graphicx}

\newcommand{\intall}{\int_{-\infty}^{\infty}}
\newcommand{\ket}[1]{|#1\rangle}

\newcommand{\bra}[1]{\langle#1|}

\newcommand{\avg}[1]{\langle#1\rangle}

\newcommand{\bk}[1]{\left(#1\right)}
\newcommand{\Bk}[1]{\left[#1\right]}

\newcommand{\trace}{\operatorname{tr}}

\begin{document}

\title{A Bayesian quasi-probability approach
to inferring the past of quantum observables}

\author{Mankei Tsang}
\email{eletmk@nus.edu.sg}
\affiliation{Department of Electrical and Computer Engineering,
  National University of Singapore, 4 Engineering Drive 3, Singapore
  117583}

\affiliation{Department of Physics, National University of Singapore,
  2 Science Drive 3, Singapore 117551}

\date{\today}


\begin{abstract}
  I describe a method of inferring the past of quantum observables
  given the initial state and the subsequent measurement results using
  Wigner quasi-probability representations. The method is proved to be
  compatible with logic for large subclasses of quantum systems,
  including those that involve incompatible observables and can still
  exhibit some quantum features, such as the uncertainty relation,
  measurement backaction, and entanglement. 
\end{abstract}

\maketitle
\section{Introduction}
Applying classical reasoning to quantum systems often leads to
paradoxes. Many of the paradoxes involve inferring the past of a
quantum observable given initial conditions and subsequent measurement
outcomes, such as the path taken by a quantum particle in an
interferometer \cite{wheeler,peres94,griffiths,*[] [{, and references
    therein.}]  aharonov_rohrlich}. The weak value proposed by
Aharonov, Albert, and Vaidman (AAV) \cite{aav} has been widely used
experimentally to provide such an inference
\cite{aharonov_rohrlich,*[] [{, and references therein.}] shikano,*[]
  [{, and references therein.}]  cho,groen,vaidman13,danan,campagne},
but it often produces results that defy common sense; for example, a
weak value can go well outside of the spectrum of an observable and
even become complex.

It is often claimed that anomalous weak values are signatures of the
non-classicality of a quantum system
\cite{aharonov_rohrlich,shikano,cho}. A recent work by Ferrie and
Combes \cite{ferrie_combes_coin} shows, however, that an estimator
analogous to the weak value can also produce non-sense when applied to
a classical coin. If the weak value does not even make sense
classically, should we expect it to be any more meaningful for quantum
systems?


Here I describe an alternative method of inferring the past of a
quantum system based on quasi-probability representations. The method
was first proposed in Refs.~\cite{smooth_pra1,smooth_pra2} and called
quantum smoothing, but the focus there was the estimation of classical
signals coupled to quantum systems and there was no general proof of
its compatibility with logic. The main purpose of this paper is to
prove that, through the celebrated Cox's theorem \cite{cox,jaynes} and
Wigner functions \cite{ferrie_rpp}, quantum smoothing is compatible
with logic for large subclasses of quantum systems, including those
that involve incompatible observables and can still exhibit some
quantum features, such as the uncertainty relation, measurement
backaction, and entanglement. This general proof is the key feature of
the proposed method and a significant improvement over previous
approaches \cite{griffiths,aharonov_rohrlich}, the logicality of which
is less obvious when incompatible observables are concerned. By
applying quantum smoothing to a qubit and the AAV gedanken experiment,
I also demonstrate how my method can avoid some of the
counter-intuitive issues associated with the weak value.

\section{Logical Inference}
\subsection{Classical}
Consider first the classical inference problem. Cox's theorem
\cite{cox,jaynes} states that the only consistent method of assigning
plausibilities to propositions is equivalent to the laws of
probability, and the Bayes theorem in particular, if one assumes a set
of desiderata that can be regarded as an extension of Aristotelian
logic and summarized as follows \cite{jaynes}:
\begin{enumerate}
\item Representation of degrees of plausibility by real numbers.
\item Qualitative correspondence with common sense.
\item Consistency.
\end{enumerate}
In the rest of the paper I shall take the view that Cox's desiderata
and logic are equivalent notions. Readers who wish to challenge this
definition of logic are urged to read Refs.~\cite{cox,jaynes}, and
until they have come up with a better definition, Cox's theorem
remains the most rigorous logical foundation available for statistical
inference.

\begin{figure}[htbp]
\centerline{\includegraphics[width=0.45\textwidth]{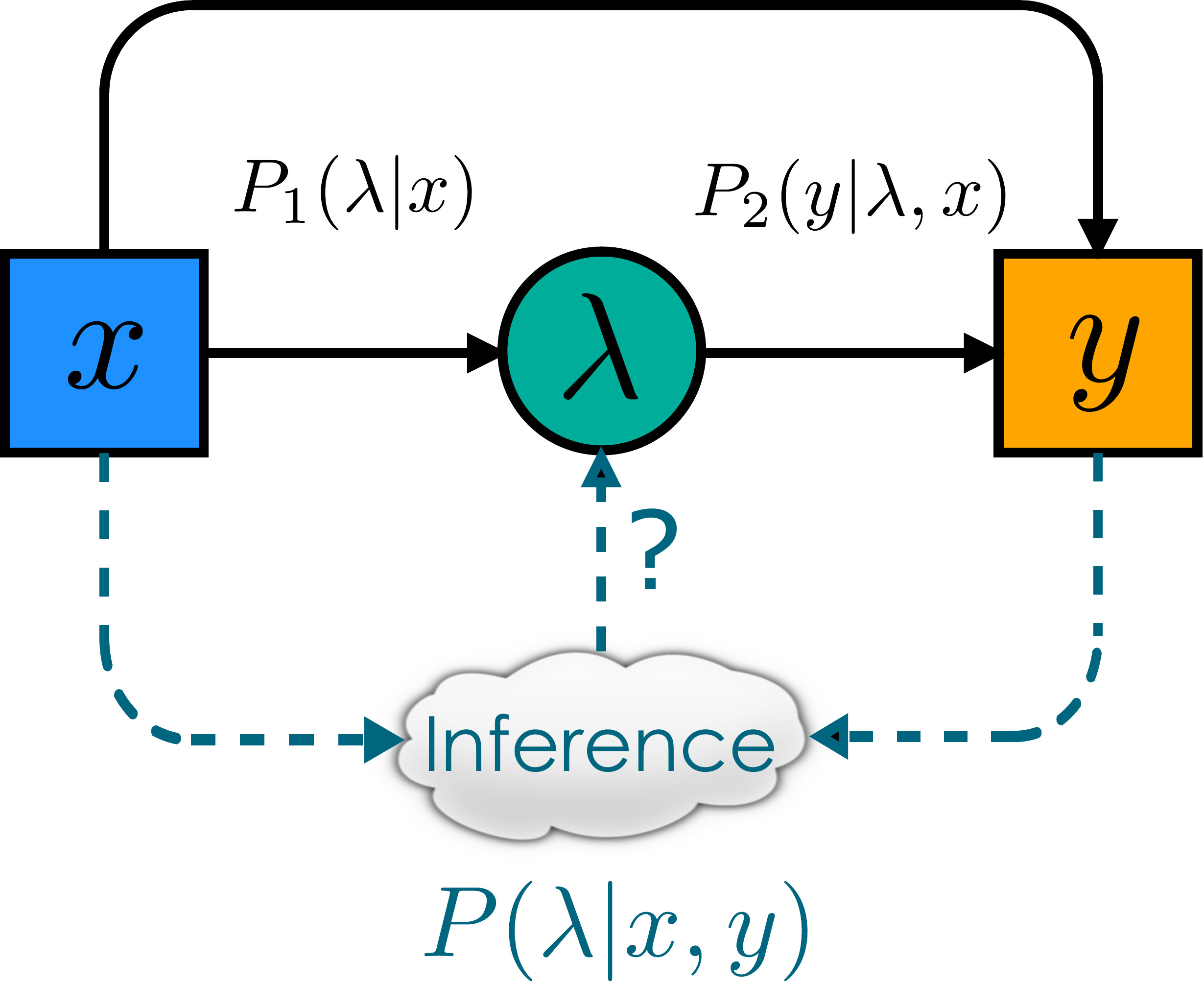}}
\caption{A flowchart depicting the inference problem. Solid arrows
  denote the direction of causality.  The problem of statistical
  inference is to infer $\lambda$ from $x$ and $y$ using the given
  $P_1(\lambda|x)$ and $P_2(y|\lambda,x)$.}
\label{smooth}
\end{figure}

To be specific, consider the inference problem depicted in
Fig.~\ref{smooth}. Let $x$ be a parameter that represents facts known
at an initial time, $y$ be another parameter that represents facts
known at a final time, and $\lambda$ be the hidden variable whose
value at the intermediate time is to be inferred. Suppose that the
predictive probability function $P_1(\lambda|x)$ and the retrodictive
likelihood function $P_2(y|\lambda,x)$ are given. This is a reasonable
assumption, as causality implies that $\lambda$ should depend on $x$,
$y$ should depend on both $\lambda$ and $x$, and noise should
introduce uncertainties to the dependencies.  The posterior
probability function of $\lambda$ conditioned on $x$ and $y$ is
\begin{align}
P(\lambda|x,y) &= \frac{P_2(y|\lambda,x)P_1(\lambda|x)}
{\sum_\lambda P_2(y|\lambda,x)P_1(\lambda|x)},
\end{align}
which is the logical assignment of plausibilities to values of
$\lambda$ given known facts $x$ and $y$.  For example, the most likely
$\lambda$ can be determined from $P(\lambda|x,y)$ by finding the
$\lambda$ that maximizes the posterior and is called the maximum
\textit{a posteriori} (MAP) estimate. In accordance with the
terminology in estimation theory \cite{smooth_pra1,smooth_pra2,simon},
I refer to the inference of a hidden variable at the intermediate time
as smoothing.

\subsection{Quantum}
I now ask how one can logically infer the past of a quantum
system. Suppose that the initial quantum state given prior facts
represented by $x$ is $\rho_x$, and the positive operator-valued
measure (POVM) for the measurement at the final time with outcome $y$
is $E(y|x)$, which, for generality, can be adaptive and depend on $x$.
Born's rule gives
\begin{align}
P(y|x) &= \trace E(y|x)\rho_x.
\end{align}
To infer logically the value of quantum observables, denoted by
$\lambda$, at the intermediate time is to assign a non-negative
posterior probability function to their possible values through the
Bayes theorem. Define a pair of maps that transform $\rho_x$ and
$E(y|x)$ to quasi-probability functions of $\lambda$:
\begin{align}
\mathcal W_1 \rho_x &= W_1(\lambda|x),
\\
\mathcal W_2 E(y|x) &= W_2(y|\lambda,x),
\end{align}
and require that they obey Born's rule in the following way:
\begin{align}
P(y|x) &= \trace E(y|x)\rho_x = \sum_\lambda W_2(y|\lambda,x)W_1(\lambda|x).
\label{born2}
\end{align}
If $W_2(y|\lambda,x)$ and $W_1(\lambda|x)$ are both non-negative, the
posterior function given by
\begin{align}
W(\lambda|x,y) &= \frac{W_2(y|\lambda,x)W_1(\lambda|x)}{P(y|x)}
\label{posterior}
\end{align}
is also non-negative and compatible with the laws of probability and
Cox's desiderata. Hence, the logicality of quantum smoothing and
the non-negativity of quasi-probability representations are equivalent
notions.

If $W_1(\lambda|x)$ and $W_2(y|\lambda,x)$ are restricted to be
non-negative, it is known that no single pair of maps $\mathcal W_1$
and $\mathcal W_2$ can explain all predictions of quantum mechanics
\cite{ferrie,spekkens}.  This is known as contextuality; without this
property, quantum mechanics would be equivalent to a classical hidden
variable model. The key point here is that many large and important
subclasses of quantum systems, such as the odd-dimensional stabilizer
quantum computation model \cite{gross_hudson,gross07} and the linear
Gaussian model for canonical observables
\cite{hillery,braunstein_rmp,bartlett2012}, turn out to be perfectly
described by non-negative $W_1(\lambda|x)$ and $W_2(y|\lambda,x)$ if
they are chosen to be the appropriate discrete or continuous Wigner
representations of $\rho_x$ and $E(y|x)$. These models generally
involve incompatible observables and can still exhibit some quantum
features, such as the uncertainty relation, measurement backaction,
and entanglement.

To prove the logicality of quantum smoothing for the aforementioned
subclasses explicitly, I write Born's rule in the following way:
\begin{align}
P(y|x) &= \trace E(y|x,t_j)\rho_x(t_j)
\end{align}
for any time $t_j, j = 0,1,\dots,J$, where 
\begin{align}
\rho_x(t_j) &\equiv \mathcal U_x(t_{j})\dots \mathcal U_x(t_2)
\mathcal U_x(t_1)\rho_x(t_0),
\\
E(y|x,t_j) &\equiv \mathcal U_x^*(t_{j+1})\dots \mathcal U_x^*(t_{J-1})\mathcal U_x^*(t_{J})E(y|x,t_J),
\end{align}
$\rho_x(t_0)$ is the initial density operator, $E(y|x,t_J)$ is the
final POVM, and $\mathcal U_x(t_j)$ denotes a unitary operation with
unitary operator $U_x(t_j)$:
\begin{align}
\mathcal U_x(t_j)\rho &\equiv U_x(t_j) \rho U_x^\dagger(t_j),
\\
\mathcal U_x^*(t_j)E &\equiv U_x^\dagger(t_j)E U_x(t_j).
\end{align}
The unitary operations can also model open-system evolution and
sequential and adaptive measurements if the Hilbert space is suitably
dilated \cite{kraus,wiseman_milburn}.  If I choose $\mathcal W_1$ to
be a Wigner representation of $\rho_x(t_j)$:
\begin{align}
W_1(\lambda,t_j|x) &= \mathcal W_1 \rho_x(t_j),
\end{align}
and $\mathcal W_2$ to be $\mathcal W_1$ multiplied
by a normalization factor $\mathcal N$:
\begin{align}
W_2(y|\lambda,x,t_j) &= \mathcal W_2 E(y|x,t_j) 
= \mathcal N\mathcal W_1 E(y|x,t_j),
\end{align}
a fundamental property of the Wigner representation \cite{ferrie} can
be used to give
\begin{align}
\trace E(y|x,t_j)\rho_x(t_j) 
&= \sum_\lambda W_2(y|\lambda,x,t_j) W_1(\lambda,t_j|x),
\end{align}
making the hidden-variable model consistent with Born's rule. For
continuous variables, $\sum_\lambda$ should be replaced by an
integral.

For quantum systems with odd dimensions, Gross
\cite{gross_hudson,gross07} showed that a pure state is a stabilizer
state if and only if the natural analog of the Wigner representation
for odd dimensions is non-negative. Restricting $\rho_x(t_0)$ to
stabilizer states and $\mathcal U_x(t_j)$ to Clifford operations,
which transform stabilizer states to stabilizer states,
$W_1(\lambda,t_j|x)$ is non-negative at any time. If $E(y|x,t_J)$ is
a stabilizer-state projection, $W_2(y|\lambda,x,t_j)$ is non-negative
as well, since $\mathcal U_x^*$ is also a Clifford operation if
$\mathcal U_x$ is one. Hence, the odd-dimensional stabilizer model,
which consists of stabilizer states, Clifford operations, and
stabilizer-state projections, can always be represented by
non-negative $W_1$ and $W_2$ at any time $t_j$ and permits logical
smoothing inference.

Similarly, for canonical observables, such as the continuous positions
and momenta of harmonic oscillators, it is well known that a state
with a Gaussian Wigner representation remains Gaussian if the
Hamiltonian is quadratic with respect to the canonical observables
\cite{hillery,braunstein_rmp,bartlett2012}. If $E(y|x,t_J)$ is a
projective measurement with respect to the canonical observables, and
the unitary operations are restricted to those with quadratic
Hamiltonians, $W_2(y|\lambda,x,t_j)$ is also Gaussian. The linear
Gaussian model, which consists of Gaussian states, quadratic
Hamiltonians, and canonical-observable measurements, can therefore be
represented by non-negative and Gaussian $W_1$ and $W_2$ at all times
and also permits logical smoothing inference.

For quantum systems that do not admit non-negative quasi-probability
representations, the smoothing inference necessarily violates Cox's
desiderata and can lead to paradoxes.  One example is shown
in Ref.~\cite{smooth_pra2} for Hardy's paradox \cite{hardy} using a
discrete Wigner representation.  It must be emphasized that negative
quasi-probabilities do not resolve any paradox; they simply mean that
the inference according to the hidden-variable model is illogical.
For illustrative and pedagogical purposes it is still useful to report
the quasi-probability functions for experiments, as they are
more obvious indicators of non-classicality than density matrices and
POVMs. For a more sensible estimate than the weak value, one can
choose
\begin{align}
\lambda_{x,y}^\textrm{MAP} &\equiv \arg \max_\lambda W(\lambda|x,y)
\end{align}
as the MAP quasi-estimate, which is still one of the possible values
of $\lambda$ and will never become anomalously large or complex, thus
avoiding some of the counter-intuitive features of the weak value. In
the case of Hardy's paradox, Ref.~\cite{smooth_pra2} shows that
$\lambda_{x,y}^\textrm{MAP}$ indeed reproduces the most likely paths
suggested by classical reasoning, even though they still lead to
logical contradictions.

\subsection{Weak value as an estimate}
In the weak-value approach, an additional weak measurement
\cite{davies,kraus,braginsky,wiseman_milburn} is made before the final
measurement.  To model that scenario, it is important not to confuse
the weak measurement outcome with the hidden variable to be
inferred. The weak measurement outcome can be grouped with either $x$
or $y$, and the Bayesian protocol, if it exists, will give us a
posterior distribution of $\lambda$ for any number of trials. From the
perspective of Cox's theorem, any method that deviates from the
Bayesian approach is illogical \cite{jaynes}, and if we view the weak
value simply as an inference method that combines data in a special
way to produce an estimate of the observable, it should not surprise
us that it can produce non-sense, unless it happens to agree with the
Bayesian approach. No amount of heuristic reasoning can save the weak
value from illogicality otherwise.

In the context of the quasi-probability framework, one can ask whether
the weak value corresponds to an estimate arising from a posterior
quasi-probability function.  The answer was provided by Johansen and
Luis \cite{johansen_luis}, who showed that the weak value is
equivalent to the conditional average using what they called the $S$
distributions, which can be negative or even complex. It is currently
unknown under what situation the $S$ distributions are non-negative,
so the logical foundation of the weak value remains questionable.

\subsection{Logical inference versus decision theory}
Besides the logical interpretation, one can also justify Bayesian
inference in a more utilitarian manner using decision theory
\cite{jaynes,berger}, which shows that Bayesian inference can minimize
the expected error between an estimate and the true value of a hidden
variable. In this paper, I do not attach any decision-theoretic
significance to the quantum inference, as the past of a quantum
observable is usually not available for error evaluation because of
the no-cloning theorem.  Any apparent illogicality that arises from
negative quasi-probabilities exists only in the mind and can be
attributed to the wrong model being used for a quantum system; the
experimenter only has access to prior facts and measurement outcomes,
which obey Born's rule and hence the laws of probability and logic.

This kind of mental exercise is not entirely philosophical however.
The smoothing inference provides an alternative way of thinking about
when a quantum system follows classical logic internally and is
therefore simulable by a classical computer. This question is central
not only to quantum computation and quantum simulation \cite{nielsen},
but also to the implementation of quantum estimation and control
algorithms \cite{smooth_pra1,smooth_pra2,smooth,wiseman_milburn}.

The method presented here is naturally extensible to the inference of
classical signals coupled to quantum systems, a problem studied
extensively in Refs.~\cite{smooth_pra1,smooth_pra2,smooth}. In that
case, it must be emphasized that, although the quasi-probability
functions can be used as an intermediate and often convenient step,
the end result is always consistent with both logic and decision
theory. The hybrid smoothing method is known to be optimal and
superior to conventional prediction or filtering methods for certain
quantum waveform estimation problems
\cite{twc,wheatley,yonezawa,iwasawa}, analogous to the classical case
\cite{simon,vantrees}, whereas classical parameter estimation based on
the weak-value approach often turns out to be suboptimal
\cite{knee,knee_prx,tanaka,ferrie_combes}.

\section{Discrete Wigner representation of qubits}
\begin{figure}[htbp]
\centerline{\includegraphics[width=0.45\textwidth]{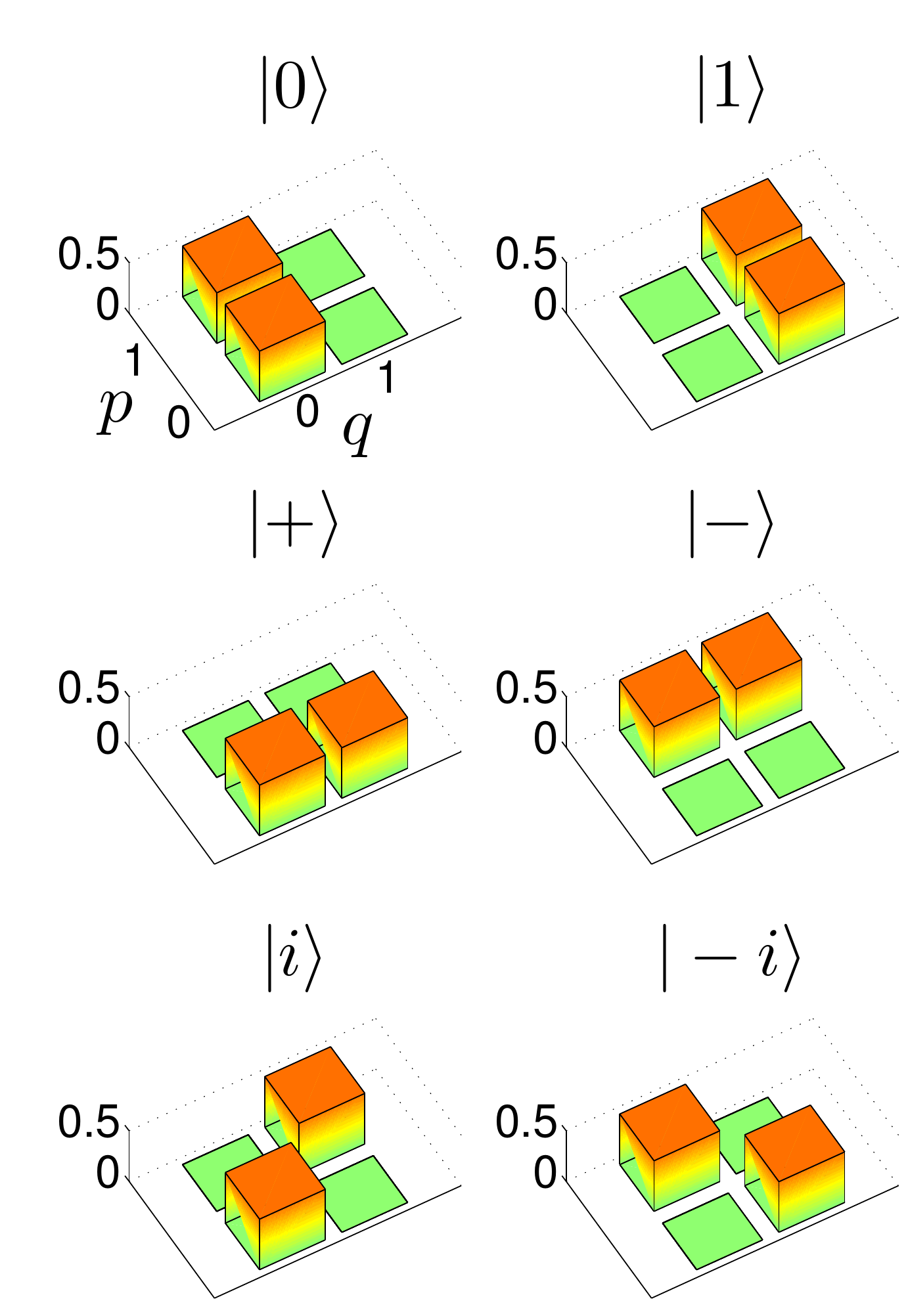}}
\caption{A discrete Wigner representation of some qubit states.}
\label{wigner}
\end{figure}

Odd-dimensional systems possess a natural and unique definition of the
discrete Wigner function \cite{gross_hudson,gross07}, but many equally
qualified definitions can exist for even dimensions \cite{gibbons}.
Here I consider the ones proposed by Feynman \cite{feynman_wigner} and
Wootters and co-workers \cite{wootters_ap,gibbons,wootters_picture}.
Consider first a qubit, which can describe the path of a quantum
particle in a two-arm interferometer, the polarization of a photon, or
the spin of an electron for example.  Define the observables of
interest as
\begin{align}
\hat q &\equiv \frac{1}{2}\bk{I-\sigma_Z},
&
\hat p &\equiv \frac{1}{2}\bk{I-\sigma_X},
&
\hat r &\equiv \frac{1}{2}\bk{I-\sigma_Y},
\end{align}
where $\sigma_Z$, $\sigma_X$, and $\sigma_Y$ are Pauli matrices and
$I$ is the identity matrix.  Define eigenstates of $\hat q$, $\hat p$,
and $\hat r$ as $\hat q\ket{0} = 0\ket{0}$, $\hat q\ket{1} = \ket{1}$,
$\hat p\ket{+} = 0\ket{+}$, $\hat p\ket{-} = \ket{-}$, $\hat
r\ket{i}=0\ket{i}$, and $\hat r\ket{-i} = \ket{-i}$.  The discrete
Wigner function proposed by Feynman \cite{feynman_wigner} and Wootters
\cite{wootters_ap} with respect to the eigenvalues of $\hat q$ and
$\hat p$ is
\begin{align}
\mathcal W_1 \rho_x
&= \frac{1}{2}\trace A(q,p) \rho_x = W_1(q,p|x),
\end{align}
where
\begin{align}
&\quad A(q,p) 
\nonumber\\
&=
\frac{1}{2}\Bk{(-1)^{q} \sigma_Z+(-1)^{p} \sigma_X
+(-1)^{q+p} \sigma_Y+I}.
\end{align}
In the phase-space matrix format,
\begin{align}
W_1(q,p|x)
&\equiv
\bk{\begin{array}{cc}
W_1(0,1|x) & W_1(1, 1|x)\\
W_1(0,0|x) & W_1(1, 0|x)
\end{array}}.
\end{align}
For example,
\begin{align}
\mathcal W_1 \ket{0}\bra{0}
&= \bk{\begin{array}{cc}0.5&0\\ 0.5 &0\end{array}},
&
\mathcal W_1 \ket{1}\bra{1}
&= \bk{\begin{array}{cc}0&0.5\\0& 0.5\end{array}},
\\
\mathcal W_1\ket{+}\bra{+}
&= \bk{\begin{array}{cc}0 &0\\0.5 &0.5\end{array}},
&
\mathcal W_1\ket{-}\bra{-}
&= \bk{\begin{array}{cc}0.5 & 0.5\\0 &0\end{array}},
\\
\mathcal W_1\ket{i}\bra{i}
&= \bk{\begin{array}{cc}0 &0.5\\0.5 &0\end{array}},
&
\mathcal W_1\ket{-i}\bra{-i}
&= \bk{\begin{array}{cc}0.5 & 0\\0 &0.5\end{array}}.
\end{align}
These are plotted in Fig.~\ref{wigner}. Note that the value of the
$\hat r$ observable can also be inferred by defining
\begin{align}
r = (q+p) \operatorname{mod} 2.
\end{align}
Note also how the uncertainty relations among the three spin
components are observed in phase space.

\begin{figure*}[htbp]
\centerline{\includegraphics[width=\textwidth]{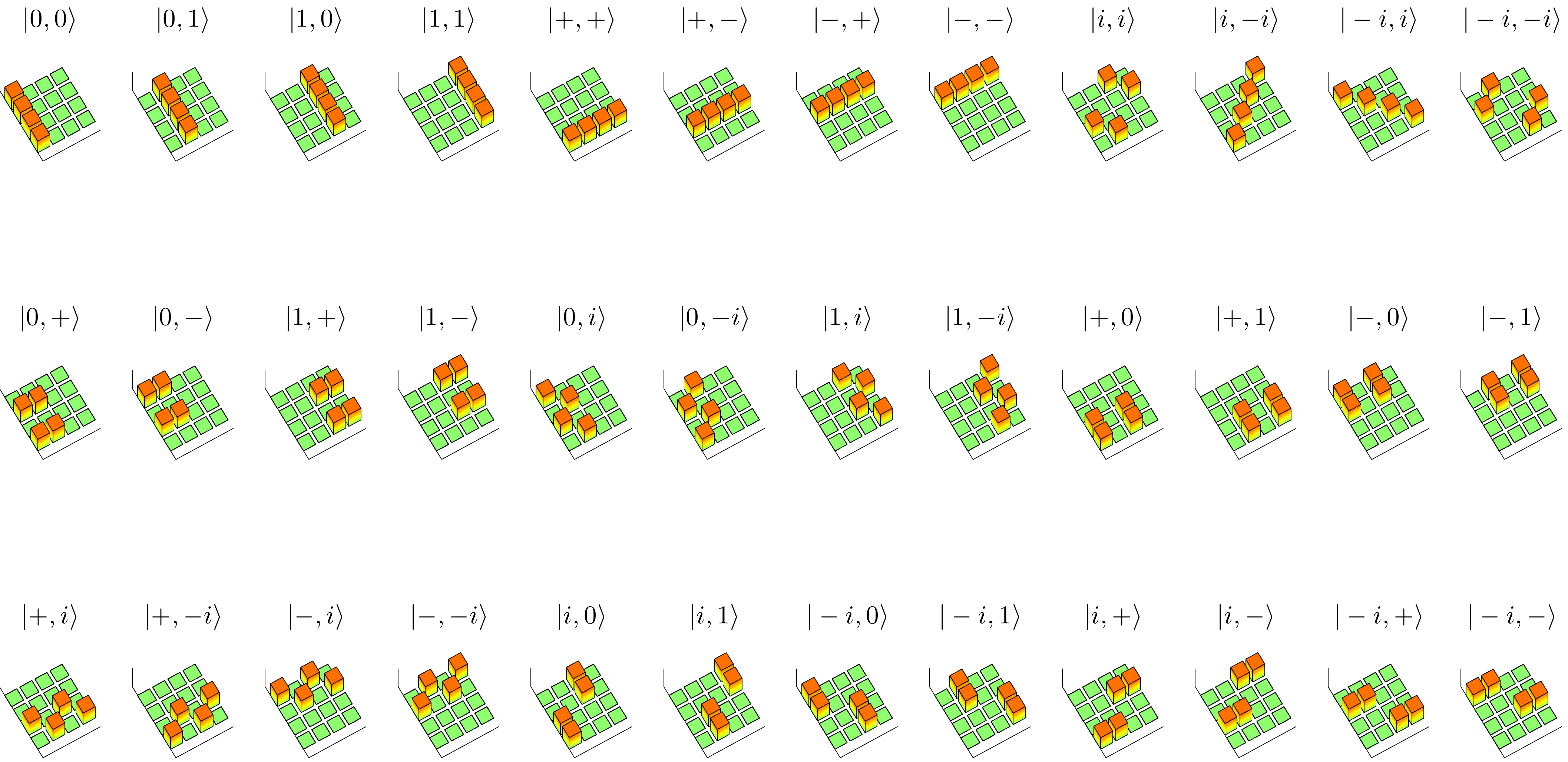}}
\caption{A discrete Wigner representation of separable two-qubit
  states.  The non-zero elements are all equal to $1/4$.}
\label{two_qubits}
\end{figure*}

\begin{figure}[htbp]
\centerline{\includegraphics[width=0.45\textwidth]{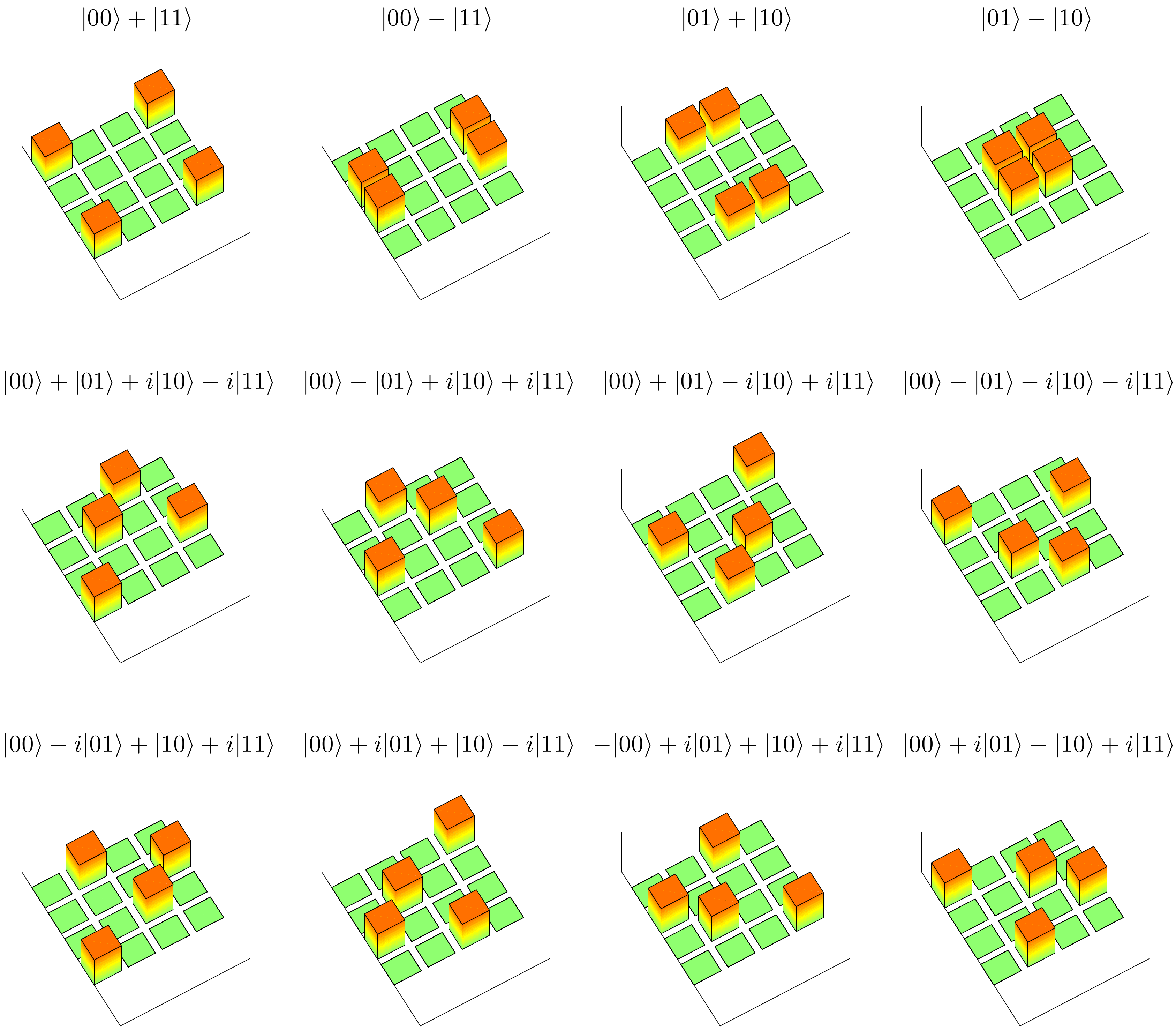}}
\caption{A discrete Wigner representation of some entangled two-qubit
  states.  Note that the states denoted in the titles are
  unnormalized for brevity.}
\label{two_qubits_entangled}
\end{figure}

For two qubits, one of the possible definitions of the Wigner
functions proposed by Gibbons \textit{et al.}\
\cite{gibbons,wootters_picture} is
\begin{align}
&\mathcal W_1\rho_x = \frac{1}{4} 
\trace A(q_{1},q_{2},p_{1},p_{2})\rho_x,
\\
&\quad A (q_{1},q_{2},p_{1},p_{2})
\nonumber\\
&= \frac{1}{2}\Bk{(-1)^{q_{1}} \sigma_{Z}+(-1)^{p_{1}} \sigma_{X}
+(-1)^{q_{}+p_{1}} \sigma_{Y}+I}\otimes
\nonumber\\&\quad
\frac{1}{2}\Bk{(-1)^{q_{2}} \sigma_{Z}+(-1)^{p_{2}} \sigma_{X}
-(-1)^{q_{2}+p_{2}} \sigma_{Y}+I}.
\label{A4}
\end{align}
Remarkably, the function stays non-negative for many separable states,
as shown in Fig.~\ref{two_qubits}, as well as some entangled states,
as shown in Fig.~\ref{two_qubits_entangled}. Unlike the case of odd
dimensions, the correspondence between non-negative Wigner
representations and the stabilizer model is not as strong for even
dimensions \cite{galvao,cormick}. For two qubits, there are 60 pure
stabilizer states \cite{aaronson04}, but the Wigner function
considered here is non-negative for only 48 of them, as shown in
Figs.~\ref{two_qubits} and \ref{two_qubits_entangled}, and becomes
negative for the remaining 12.

For the measurement, the map on the POVM can be chosen as
\begin{align}
\mathcal W_2 E(y|x)
&= N\mathcal W_1 E(y|x),
\end{align}
with $N$ being the dimension, such that a fundamental property of the
Wigner representation \cite{wootters_ap,gibbons} makes the
hidden-variable model agree with Born's rule according to
Eq.~(\ref{born2}). Figs.~\ref{wigner}, \ref{two_qubits}, and
\ref{two_qubits_entangled} then also depict the Wigner representations
of projective measurements, normalization notwithstanding. Systems with
initial states, state transitions, and measurements within this set of
states with non-negative Wigner representations naturally permit
logical smoothing inference.

\section{Logical smoothing and complex weak value}
It is not difficult to construct examples where the weak value does
not make sense, while the Wigner functions provide a logical path for
smoothing inference. Consider an initial state given by
\begin{align}
\rho_{x=0} &= \ket{0}\bra{0},
\\
\rho_{x=1} &= \ket{1}\bra{1},
\\
W_1(q,p|0) &= 
\bk{\begin{array}{cc}0.5&0\\ 0.5 &0\end{array}},
\\
W_1(q,p|1) &= 
\bk{\begin{array}{cc}0&0.5\\ 0 &0.5\end{array}},
\end{align}
and a final $\hat r$ measurement given by
\begin{align}
E(y = i) &= \ket{i}\bra{i},
\\
E(y = -i) &= \ket{-i}\bra{-i},
\\
W_2(i|q,p) &= 
\bk{\begin{array}{cc}0&1\\ 1 &0\end{array}},
\\
W_2(-i|q,p) &= 
\bk{\begin{array}{cc}1&0\\ 0 &1\end{array}}.
\end{align}
$W_1$ and $W_2$ are consistent with Born's rule and remain
non-negative for all $x$ and $y$, enabling logical inference of the
past values of $q$, $p$, and $r =
(q+p) \operatorname{mod} 2$. Suppose $x = 0$ and $y =
i$. The smoothing quasi-probability function becomes
\begin{align}
W(q,p|x=0,y=i)
&= \bk{\begin{array}{cc}0&0\\ 1 &0\end{array}},
\end{align}
which infers with certainty that $q = p = r = 0$.  This result
concerning the three incompatible observables follows from a logical
inference protocol but seems to be at odds with the uncertainty
principle. The apparent conflict can be resolved by realizing that,
under the quasi-probability framework, the uncertainty relation holds
only for the predictive part $W_1$, and a smoothing inference can
violate the principle even if it is logical.  The weak value of $\hat
p$, on the other hand, is
\begin{align}
\frac{\bra{i}\hat p\ket{0}}{\avg{i|0}}&= \frac{1+i}{2},
\end{align}
which is complex and makes no sense as an estimator. One could take
the real part or the absolute value of the weak value to make it look
more reasonable, but again there is no logical foundation that
justifies such heuristic operations.

This example can also be studied using the approach of consistent
histories \cite{griffiths}. The coherence functional is the central
quantity in this approach and for the present example with $x = 0$ and
$y = i$ is defined as $C(p,p')$ and given by
\begin{align}
C(0,0) &= \trace \ket{i}\avg{i|+}\avg{+|0}
\avg{0|+}\avg{+|i}\bra{i} = \frac{1}{4},
\\
C(0,1) &= 
\trace \ket{i}\avg{i|+}\avg{+|0}\avg{0|-}\avg{-|i}\bra{i} = -\frac{i}{4},
\\
C(1,0) &= 
\trace \ket{i}\avg{i|-}\avg{-|0}\avg{0|+}\avg{+|i}\bra{i} = \frac{i}{4},
\\
C(1,1) &= 
\trace \ket{i}\avg{i|-}\avg{-|0}\avg{0|-}\avg{-|i}\bra{i} = \frac{1}{4}.
\end{align}
Although the off-diagonal components $C(0,1)$ and $C(1,0)$ are not
zero, they are imaginary and still satisfy the ``weak consistency
condition'' defined in Ref.~\cite{griffiths}. The general relation
between the consistent histories approach and the quasi-probability
approach here is an interesting open problem but beyond the scope of
this paper.

\section{The Aharonov-Albert-Vaidman gedanken experiment}

\begin{figure}[htbp]
\centerline{\includegraphics[width=0.45\textwidth]{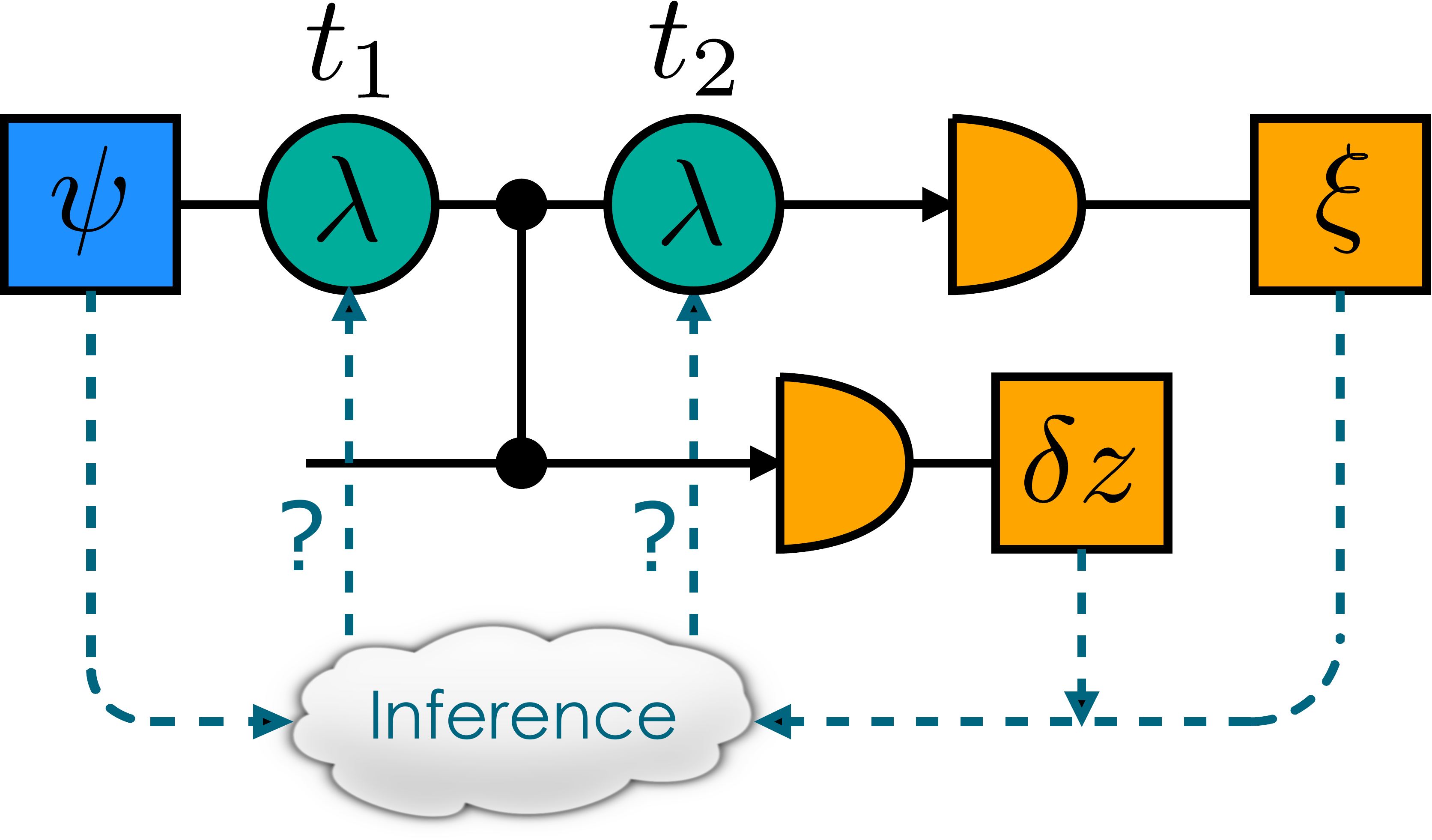}}
\caption{A flowchart (quantum circuit diagram) depicting the
  Aharonov-Albert-Vaidman gedanken experiment. Starting with an
  initial state $\psi$, inference of the hidden variable $\lambda$ at
  times $t_1$ and $t_2$ is performed using a weak $q$ measurement with
  outcome $\delta z$ and a final projective $p$ measurement with
  outcome $\xi$.}
\label{aav}
\end{figure}

To demonstrate the limitations of the quasi-probability approach,
consider the original experiment proposed by AAV \cite{aav}, as
depicted in Fig.~\ref{aav}.  A spin-1/2 particle is known to be in a
pure state $\ket\psi$ at time $t_1$, A weak $\hat q$ measurement is
then performed and can be modeled by the Kraus operator
\cite{kraus,wiseman_milburn}:
\begin{align}
K(\delta z) &=
\frac{1}{(2\pi \delta t)^{1/4}}
\exp\Bk{-\frac{(\delta z-\hat q\delta t)^2}{4\delta t}}
\\
&\approx
\frac{1}{(2\pi \delta t)^{1/4}}
\exp\bk{-\frac{\delta z^2}{4\delta t}}
\bk{1+\frac{\delta z}{2}\hat q-\frac{\delta t}{8}\hat q^2},
\end{align}
where $\delta t$ characterizes the measurement strength, assumed to be
unitless and $\ll 1$. The time after the weak measurement and before
the final measurement is denoted as $t_2$.  The final measurement is
an $p$ projection:
\begin{align}
E(\xi = +) &= \ket{+}\bra{+},
\\
E(\xi = -) &= \ket{-}\bra{-},
\end{align}
such that Born's rule is given by
\begin{align}
P(\xi,\delta z|\psi)
&= \trace \ket\xi\bra\xi K(\delta z)\ket{\psi}
\bra{\psi} K^\dagger(\delta z).
\label{born3}
\end{align}

It is important here not to confuse the measurement outcome $\delta z$
with the hidden observable it is measuring. The weak-value approach
calculates the average of $\delta z/\delta t$ for many trials and
claims that it is an estimate of $q$, but such an approach cannot be
justified unless it happens to agree with a Bayesian estimate. The
approach would work for an analogous classical problem because there
exists a likelihood function describing the weak measurement such that
averaging the outcomes over many trials is akin to evaluating $\intall
d(\delta z) P_2(\delta z|q) (\delta z/\delta t) = q$, which gives the
true $q$.  For the quantum problem, however, it is not obvious how
such a non-negative likelihood function can be defined to justify the
weak value as an estimate, unless the Kraus operator $K(\delta z)$
happens to commute with all the other operators in the problem.

\begin{figure}[htbp]
\centerline{\includegraphics[width=0.45\textwidth]{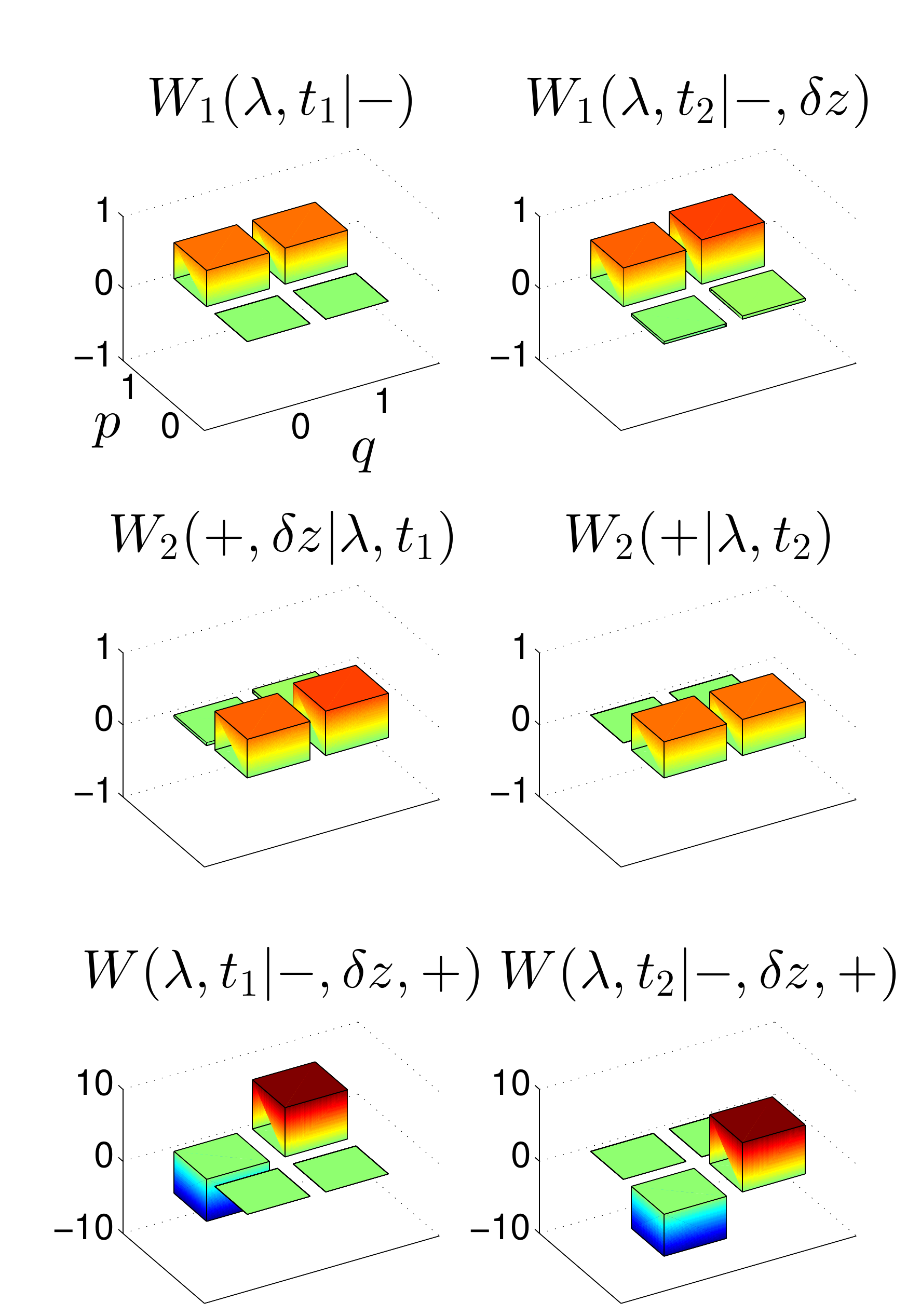}}
\caption{The predictive (1st row), retrodictive (2nd row), and
  smoothing (3rd row) quasi-probability functions before and after the
  weak measurement for $\delta t = 0.1$ and $\delta z = \sqrt{0.1}$.
  Note that the $W_1(\lambda,t_2|-,\delta z)$, $W_2(+,\delta
  z|\lambda,t_1)$, and $W_2(+|\lambda,t_2)$ plotted here are
  unnormalized. All the functions must be non-negative for the
  inference to conform to logic; the negativity shown here is a
  signature of non-classicality.}
\label{qsmooth}
\end{figure}

Let's focus on the case $\ket\psi = \ket-$ with the final result $\xi
= +$, which gives an infinite weak value \cite{aav}.  The $\xi = +$
final result is possible because of the small backaction noise
introduced by the weak measurement to $p$.  Consider the predictive
Wigner functions $W_1(\lambda,t_1|\psi)$ and
$W_1(\lambda,t_1|\psi,\delta z)$ before and after the weak
measurement:
\begin{align}
W_1(\lambda,t_1|-) &\equiv
\mathcal W_1\ket{-}\bra{-}  = 
\frac{1}{2}
\bk{\begin{array}{cc}1 & 1\\ 0& 0\end{array}},
\\
W_1(\lambda,t_2|-,\delta z)
&\equiv
\mathcal W_1
\frac{K(\delta z)\ket{-}\bra{-}K^\dagger(\delta z)}
{\trace(\textrm{numerator})}
\nonumber\\
&\propto
\bk{\begin{array}{cc}1 & 1\\ 0& 0\end{array}}
+\frac{\delta z}{2}
\bk{\begin{array}{cc}0.5 & 1.5\\ -0.5& 0.5\end{array}}
\nonumber\\
&\quad +\frac{\delta t}{8}
\bk{\begin{array}{cc}-0.5&-0.5\\0.5&0.5\end{array}}
,
\end{align}
and the following retrodictive quasi-likelihood functions:
\begin{align}
W_2(+,\delta z|\lambda,t_1)
&\equiv \mathcal W_2 K^\dagger(\delta z)\ket{+}\bra{+}K(\delta z)
\nonumber\\
&\propto \bk{\begin{array}{cc}0&0\\
1&1\end{array}}
+\frac{\delta z}{2}\bk{\begin{array}{cc}-0.5&0.5\\0.5&1.5\end{array}}
\nonumber\\&\quad
+\frac{\delta t}{8}
\bk{\begin{array}{cc}0.5&0.5\\-0.5&-0.5\end{array}},
\\
W_2(+|\lambda,t_2)&\equiv
\mathcal W_2 \ket{+}\bra{+}
= \bk{\begin{array}{cc}0&0\\1&1\end{array}}.
\end{align}
The smoothing quasi-probability functions
$W(\lambda,t_{1,2}|\psi,\delta z,\xi)$ at times $t_1$ and $t_2$ are
then
\begin{align}
W(\lambda,t_1|-,\delta z,+)
&\approx 
\bk{\begin{array}{cc}1/2-2\delta z/\delta t&
1/2+2\delta z/\delta t\\0&0\end{array}},
\\
W(\lambda,t_2|-,\delta z,+)
&\approx
\bk{\begin{array}{cc}0&0\\
1/2-2\delta z/\delta t&
1/2+2\delta z/\delta t\end{array}}.
\end{align}
All the predictive, retrodictive, and smoothing quasi-probability
functions are plotted in Fig.~\ref{qsmooth}.

There are a few interesting observations to be made:
\begin{itemize}
\item Both $W_1(\lambda,t_2|-,\delta z)$ and $W_2(+,\delta
  z|\lambda,t_1)$ become negative when $\delta z \neq 0$, rendering
  the inference illogical. The smoothing quasi-probability functions
  still agree with common sense about $p$ however, as they
  infer that $p$ must be $1$ at $t_1$ because of the initial
  state and $0$ at $t_2$ because of the final measurement outcome.
\item The marginal smoothing distributions with respect to $q$
  can be written as
\begin{align}
W(q,t_1|-,\delta z,+)
&\approx
W(q,t_2|-,\delta z,+)
\nonumber\\
&\approx
\bk{\frac{1}{2}-2\frac{\delta z}{\delta t}
\quad
\frac{1}{2}+2\frac{\delta z}{\delta t}}.
\end{align} 
Even though the negativity of Wigner functions is limited, one sees
here that the smoothing quasi-probabilities that result from them can
have arbitrarily negative values if $|\delta z/\delta t| > 1/4$.

\item The MAP quasi-estimate of $q$ is
\begin{align}
q^\textrm{MAP} &= 
\left\{\begin{array}{cc}
0, &\delta z < 0,\\
1, &\delta z > 0,\\
\textrm{ambiguous}, & \delta z = 0.
\end{array}\right.
\end{align}
This makes sense, as the outcome $\delta z$ of a $q$ measurement,
however noisy, should constitute evidence that should persuade the
observer one way or the other. Because of the negative
quasi-probabilities, the quasi-estimate should not be taken seriously
as a logical estimator, but it is at least more sensible
than the infinite weak value or, say, the naive conditional average:
\begin{align}
\bar q \equiv 
\sum_{q}q W(q|-,\delta z,+)
&\approx \frac{1}{2}+2\frac{\delta z}{\delta t}.
\end{align}
$\bar q$ exceeds the range $[0,1]$ when a smoothing
quasi-probability becomes negative, that is, when $|\delta z/\delta t|
> 1/4$. Since $\delta z \in (-\infty,\infty)$ and is typically on the
order of $\sqrt{\delta t}$, the magnitude of $\bar q$ can
become extremely large, like the weak value. This anolamy is simply
another manifestation of the illogicality that arises from the
negative quasi-probabilities.
\end{itemize}

\section{Other related work}
The Bayesian inference of an intermediate quantum projective
measurement outcome was first considered by Watanabe \cite{watanabe}
and Aharonov, Bergmann, and Lebowitz \cite{abl}. Yanagisawa first
introduced the term quantum smoothing and applied it to quantum
non-demolition (QND) observables, which are compatible observables in
the Heisenberg picture
\cite{yanagisawa}. Refs.~\cite{smooth_pra1,smooth_pra2,smooth} focus
on the smoothing inference of classical stochastic waveforms coupled
to quantum systems under continuous measurements, while more recent
papers by Dressel, Agarwal, and Jordan \cite{dressel,dressel12} and
Gammelmark, Julsgaard, and M\o{}lmer \cite{gammelmark2013} extend the
theory to the inference of weak measurement outcomes. These results
can be regarded as sharing the same foundation, as projective or weak
measurement outcomes and classical random variables can all be modeled
as QND observables in a suitably dilated Hilbert space
\cite{belavkin_qnd}.  A collection of QND observables that commute
with each other at all times of interest are called a
quantum-mechanics-free subsystem in Ref.~\cite{qmfs} to emphasize that
they have no quantum feature. It is easy to show that QND observables
always have consistent histories \cite{griffiths}.

The inference of QND observables is always compatible with decision
theory, since they can be measured without any backaction and compared
with the estimates for error evaluation, but the theory is not as
general as the one proposed here and in
Refs.~\cite{smooth_pra1,smooth_pra2}, as quasi-probability
distributions generally involve incompatible observables. Another
recent work by Chantasri, Dressel, and Jordan \cite{chantasri} also
proposes a phase-space approach to the quantum smoothing problem, but
their method is based on path integrals and its connection with more
well known and useful quasi-probability functions is
unclear.

\section{Conclusion}
The Wigner representations are currently some of the best tools for
finding classical models of quantum systems
\cite{ferrie_rpp,testing_quantum}, and negative Wigner
quasi-probability is known to be a necessary resource for quantum
computation \cite{veitch2012,veitch2013,mari12}. By equating the
logicality of quantum smoothing and the non-negativity of
quasi-probability representations, I have also made a connection
between the quantum smoothing inference problem and the notion of
contextuality \cite{ferrie,spekkens}. These connections suggest that
the smoothing method based on logical inference and Wigner functions
is a pretty good, if not the best, attempt at reconciling logic and
quantum mechanics when one tries to infer the past of a quantum
system, and further progress along these lines will benefit multiple
areas of quantum information processing and quantum foundations.

\section*{Acknowledgments}
Inspiring and helpful discussions with Joshua Combes, Christopher
Ferrie, Justin Dressel, and George Knee are gratefully
acknowledged. This work is supported by the Singapore National
Research Foundation under NRF Grant No. NRF-NRFF2011-07.

\bibliography{research}

\end{document}